\pdfoutput=1
\documentclass[sigconf]{acmart}

\acmConference[Conference'19]{}{April, 2019}{Montreal, Canada}
\acmYear{2019}
\copyrightyear{2019}

\usepackage{booktabs} 

\newcommand{\Hair}{\ifmmode\mskip1mu\else\kern0.08em\fi}

\setcopyright{rightsretained}

\usepackage{graphicx}
\graphicspath{{./figures/}}
\usepackage{float}

\let\sigproof\proof\let\proof\relax
\let\sigendproof\endproof\let\endproof\relax

\usepackage{amsthm}

\let\proof\sigproof
\let\endproof\sigendproof

\newtheoremstyle{sig}
  {}
  {}
  {\itshape}
  {}
  {\scshape}
  {.}
  {.5em}
  {#1 #2\thmnote{\quad(#3)}}

\theoremstyle{sig}
\newtheorem{dfn}{Definition}
\newtheorem{cond}{Condition}

\begin{document}
\title{A Multilevel Cybersecurity and Safety Monitor for Embedded Cyber-Physical Systems}

\author{Smitha Gautham}
\affiliation{%
   \institution{Virginia Commonwealth University}
}

\email{gauthamsm@vcu.edu}

\author{Georgios Bakirtzis}
\affiliation{%
   \institution{Virginia Commonwealth University}
}

\email{bakirtzisg@ieee.org}

\author{Matthew T. Leccadito}
\affiliation{%
   \institution{Virginia Commonwealth University}
}

\email{leccaditomt@vcu.edu}

\author{Robert H. Klenke}
\affiliation{%
   \institution{Virginia Commonwealth University}
}

\email{rhklenke@vcu.edu}

\author{Carl R. Elks}
\affiliation{%
   \institution{Virginia Commonwealth University}
}

\email{crelks@vcu.edu}

\begin{abstract}
Cyber-physical systems (\textsc{cps}) are composed 
of various embedded subsystems
and require specialized software, firmware, and hardware 
to coordinate with the rest of the system.
These multiple levels of integration
expose attack surfaces which can be susceptible
to attack vectors that require novel architectural methods
to effectively secure against.
We present a multilevel hierarchical monitor architecture cybersecurity approach
applied to a flight control system.
However, the principles present in this paper apply
to any \textsc{cps}.
Additionally, the real-time nature of these monitors allow
for adaptable security, meaning that they mitigate against possible classes
of attacks online.
This results in an appealing bolt-on solution that is independent
of different system designs.
Consequently, employing such monitors leads 
to strengthened system resiliency
and dependability of safety-critical \textsc{cps}.
\end{abstract}



\maketitle
\renewcommand{\shortauthors}{S. Gautham, G. Bakirtzis, M.T. Leccadito, R.H. Klenke, C.R. Elks}
\section{Introduction}
Ensuring the safety and security of high integrity \textsc{cps} applications is a difficult task.
Rigorous verification and design assurance strategies are especially important
for safety-critical embedded real-time systems where the significance of failure
or malicious exploitation systems may result in far-reaching consequences
to human life, environment, and financial loss.
In these situations, design assurance standards
and regulatory guidelines, for example, \textsc{iec} 61508, \textsc{do-254}, and \textsc{iso}-26262,
are used to provide high levels of assurance evidence
to confirm that the systems's unsafe or security failures are mitigated
to ``as low as reasonably achievable'' standard.
Furthermore, the complexity of these systems makes it increasingly difficult
to assure security through traditional perimeter-based security approaches.
Because design assurance methods are not 100\% foolproof, employing additional safety
and security measures at runtime have been used for some time~\cite{watterson:2007}.

In recent years, runtime verification methods have become more prevalent
and mature as a means to augment the safety and security of systems~\cite{kane:2015,moosbrugger:2017}.
Runtime verification tries to bridge the gap between design-time safety assurance methods
and traditional safety testings.
Additionally, runtime verification addresses the shortfalls
of design-time verification
and testing in a way that is complementary to both verification methods.

While runtime verification and monitors are not unique 
to \textsc{cps}, they can provide significant advantages in this domain. 
This is because \textsc{cps} has a specific expected service and limited functions 
in contrast to more general personal computing systems. 
Hence, making it amenable to use monitors and applying specific constraints 
or conditions on the monitored state of the system to detect and mitigate attacks. 
Monitors in the realm of \textsc{cps} can use the physics of the system to provide resilience.
For example, when an unmanned aerial system (\textsc{uas}) is at a certain altitude it can physically only move to another gradually.
Therefore, an abrupt change in altitude value could mean an attack is taking place 
or an intrinsic fault is degrading the operation of the system.
Similarly, in the computing domain, for example, communications or execution, there is expected behavior---the normal state---which is deviated 
from can indicate a potential attacker is attempting to violate system resources. 
This is an immediate consequence of viewing system security as an emergent property.

Another important aspect of monitors is their they ability 
to log the system state at all times.
This can further inform of what consists normal and abnormal states 
for the given system based on real situational data.
Furthermore, it can facilitate offline forensics analysis
in the event of a known breach that can lead to more secure
and safer \textsc{cps} architectures getting deployed later. 
This is increasingly important in safety-critical \textsc{cps}, where attacks can lead to hazardous states, 
which can in turn lead to accidents. 
Therefore, architectural solutions---in the form of application-specific security/safety monitors---can be seen both
as necessary and cost effective.

Nonetheless, embedded real-time systems 
and specifically \textsc{cps} present particular challenges for runtime monitoring.
System internals are not easily observable (such as address and data buses),
as many device features are incorporated deep 
within complex chip packages.
Some embedded \textsc{cps} have limited computational 
and memory resources or have real-time requirements 
that must adhere to strict deadlines.
Moreover, \textsc{cps} and embedded systems are vulnerable
to security threats at multiple levels
that span both hardware peripherals and software implementations.
These include sensors, actuators, application software, firmware, 
and communication networks, to name a few.

These traits suggest that focused (limited) monitoring, 
but distributed across different architectural layers 
in a \textsc{cps} may be more effective at timely detection of attacks (or failures)
and lessen the burden of monitoring overhead; high overhead could compromise core system resources 
or affect scheduling, causing interference to the target application~\cite{lu:2007}. 
Specifically, the two issues examined in this paper are: 
(1) the development of a monitoring framework capable of sufficiently observing the internal operation of a target \textsc{cps} at multiple levels and 
(2) design and implementation of a multilevel monitoring framework to preliminary characterize and discuss the challenges of realization. 

Towards these goals, we propose and develop the multilayer hierarchical architecture for implementing \textsc{cps} security/safety monitors for safety-critical applications.

\textbf{Contributions.} \quad Our contributions are:
\begin{itemize}
    \item the presentation of design principles for building and evaluating \textsc{cps} security monitors;
    \item the construction of a general-purpose hierarchy for the implementation of \textsc{cps} security monitors; and
    \item the evaluation of multilayer hierarchical architecture by implementing a \textsc{cps} security monitor on field-programmable gate array  (\textsc{fpga}) fabric and a processor for a flight control system (\textsc{fcs}) application.
\end{itemize}

\section{Background}

A \textsc{cps} is often comprised of numerous integrated components 
and subsystems interacting and communicating with each other to satisfy system (plant) level goals.
These goals are often related to the functional performance, safety and security of the service a \textsc{cps} is providing---example being, an automobile cruise control 
will always disengage when the brake is applied. 
With the functional safety applications, failure due to the presence of attackers can lead to situations where security breaches affect safety. The important considerations while designing the multilevel monitor architecture and the questions at hand while designing are:

\begin{itemize}
    \item The characteristics and vulnerabilities of the \textsc{cps} that define the threat model at each functional level of the system.
    \item What is the impact of overhead at multiple levels?
    \item The evaluation of the benefits to the system of deploying multiple monitors rather than a single monitor; that is, what are the design issues?
    \item Is the use of multiple monitors justified at the implementation stage for detections of attack vectors?
    \item Given an architecture, how does one determine where to place monitors?
\end{itemize}

In general, runtime monitoring makes use
of an external monitor that observes the execution behavior
of a target system during runtime
while making as few as possible assumptions
about the trustworthiness
or proper functioning of the monitored computer~\cite{bateman:2005,lee:1998,havelund:2002,pasqualetti:2013,kane:2014}.
For example, we can monitor discrete representations
of continuous time variables such as sensor readings, operating system parameters, power spectral density (\textsc{psd}) signatures~\cite{jacoby:2011},
and application level variables such as control states.
In these instances, the monitor, given a set 
of information states from the \textsc{cps}, checks if the system is compliant
to a reference of acceptable behavior.
A reference of acceptable behavior implies there exists checking conditions
or detection predicates
to decide upon a notion of \emph{acceptable behavior}~\cite{jones:2014}.

Multiple monitors are necessary for timely detection
of security risks
in today's \textsc{cps} architectures.
These multiple monitors work independently
and can individually detect an attack occurring
in each layer of a \textsc{cps}.

Prior work recognizes this need for multiple monitors.
Lu et al.~\cite{lu:2015} propose an architectural approach
where the system is divided into three layers:
the execution layer, the transport layer, and the control layer
to monitor attacks against the system hardware, the communication network,
and the control policies. 
Liu et al.~\cite{liu:2017} model a \textsc{cps} by dividing the system
on the basis of available physical, communication,
and computational entities.
Further in this work various ways of ensuring security
of each of the entities and their interactions are surveyed.

The hierarchical approach proposed here transcend various layers 
and ensures hardware, informational, 
and executional integrity building 
on the notion of hierarchical architecture monitors~\cite{leccadito:2017}.

\begin{figure}[!t]
    \includegraphics[width=1.0\linewidth]{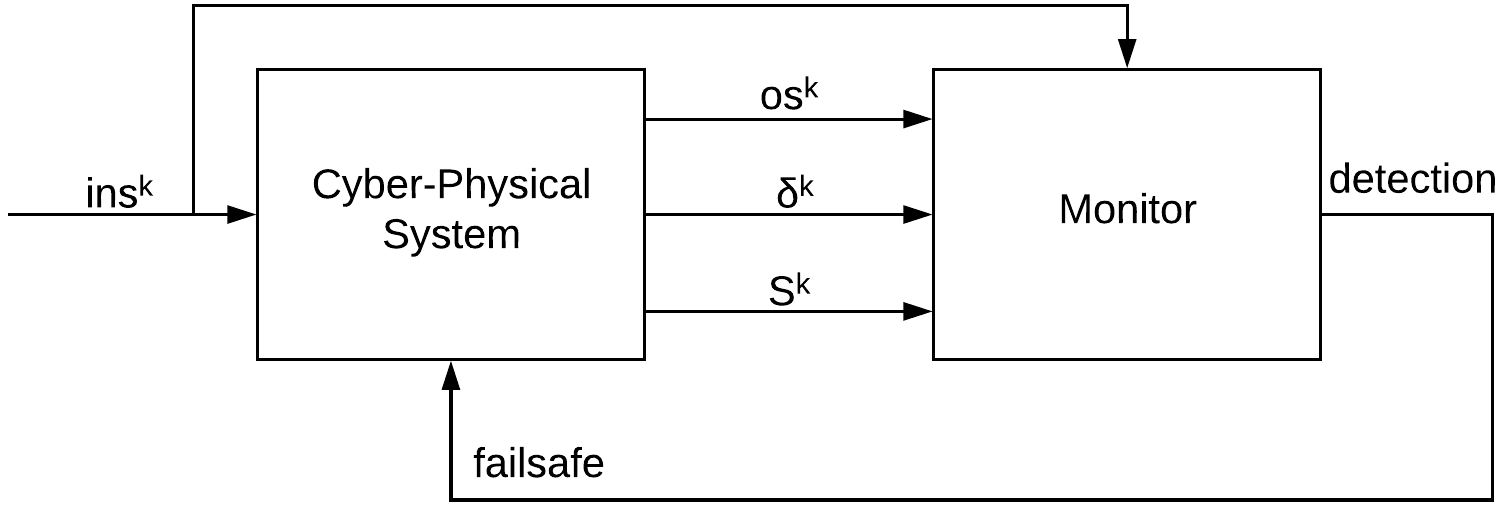}
    \caption{The single monitor setup observes the \textsc{cps} to produce a detection, at which point it attempts to mitigate the issue.}
    \label{fig:single_monitor}
\end{figure}

\section{Formal Development of Multilevel Runtime Monitoring}

\subsection{Definitions}
We develop a formal model 
of multilevel monitors 
using event calculus and graph theory.
We use that formal model
in the implementation
of the monitor's
detector function (Section~\ref{sec:imp}).

In the case of a single monitor the behavior
of the \textsc{cps} is observed and investigated
for faults and security violations by a monitor (Figure~\ref{fig:single_monitor}).
As with before we denote a monitor as \(M\).
We assume that the monitor observes \emph{stream}
of information from the target system, which includes its interactions with the environment,
and sends the a fail-safe state when an abnormal condition, for example, fault or attack, is detected.
These streams can be discrete or continuous time states, sensor data, instructions,
or other external states, for example, \textsc{gps}, time, etc.
Due to the reactive nature of \textsc{cps}, we view streams
of information from a historical perspective.
That is, from the current state of the \textsc{cps}
to states previous in time, or past time temporal observations. 

Each stream can be a sequence of inputs
or outputs to and from the \textsc{cps}
for each point in time.
We denote the \(k\)th prefix for past \(m\) instances
of an infinite stream, \(s\) as \(s^k = (s (k-m), \dots, s (k-2), s (k-1), s(k))\).
Thus, the stream at an instance \(k\) includes information
of the past \(m\) instances starting with \(s (k-m)\) and ending
at the current instance \(s(k)\).

The monitor also observes the control flow transitions \(\delta\)
of the \textsc{cps}, to keep track
of the execution flow in the design.
The transition flow
at time \(k\) \(\delta^k\) is: \(\delta^k = \left(\delta (k-m), \dots, \delta (k-2), \delta (k-1), \delta(k)\right)\).
Hence, the monitor observes the current state 
of the the \textsc{cps} as constructed
from all previous states: \(S^k_1, S^k_2, \dots, S^k_m\).
The language of the monitor is defined
by the set of monitored streams, \(\mathcal{M}_w = \left(ins^k, os^k, \delta^k, S^k_1, S^k_2, \dots, S^k_m\right)\),
where \(ins^k\) the input stream, \(os^k\) the output stream, \(\delta^k\) the execution transition data stream, and \(S^k_1, S^k_2, \dots, S^k_m\) the observed states from current time \(k\)
for the past \(m\) instances.

The role of a monitor \(M\) is to answer the proposition ``given a set of monitored streams from a process, did the proper sequence of operations occur in that process, and is the result safe and secure'' based on observed streams of sequenced events \(\mathcal{M}_w\). These events can be characterized as language or alphabet of the safety monitor. In the case where these events are compromised by security exploits or faulty behavior they may not be contained in the language and are termed as the anomalous or out-of-specification events. If there is an attack, and that attack is observable and is contained in \(\mathcal{M}_w\), then we want the monitor \(M\) to be able to detect the attack.

The monitor also observes the control flow transitions \(\delta\)
of the \textsc{cps}, to keep track
of the execution flow in the design.
The transition flow
at time \(k\), \(\delta^k\) is: \(\delta^k = \left(\delta (k-m), \dots, \delta (k-2), \delta (k-1), \delta(k)\right)\).
Hence, the monitor observes the current state 
of the the \textsc{cps} as constructed
from all previous states: \(S^k_1, S^k_2, \dots, S^k_m\).
The language of the monitor is defined
by the set of monitored streams, \(\mathcal{M}_w = \left(ins^k, os^k, \delta^k, S^k_1, S^k_2, \dots, S^k_m\right)\),
where \(ins^k\) the input stream, \(os^k\) the output stream, \(\delta^k\) the execution transition data stream, and \(S^k_1, S^k_2, \dots, S^k_m\) the observed states from current time \(k\)
for the past \(m\) instances.

We define the attack and fault detection concept more clearly below by defining monitors that recognize safety properties of the target \textsc{cps}.

\begin{dfn}[Monitor]
Following from above, a monitor, \(M\), is a distinct module
that recognizes a set of traces or streams
\(\mathcal{M}_w = \left(ins^k, os^k, \delta^k, S^k_1, S^k_2, \dots, S^k_m\right)\)
from a target \textsc{cps} whose computations can be represented as a monitorable safety language~\cite{viswanathan:2000}.
\end{dfn}

\begin{figure}[!t]
    \includegraphics[width=1.0\linewidth]{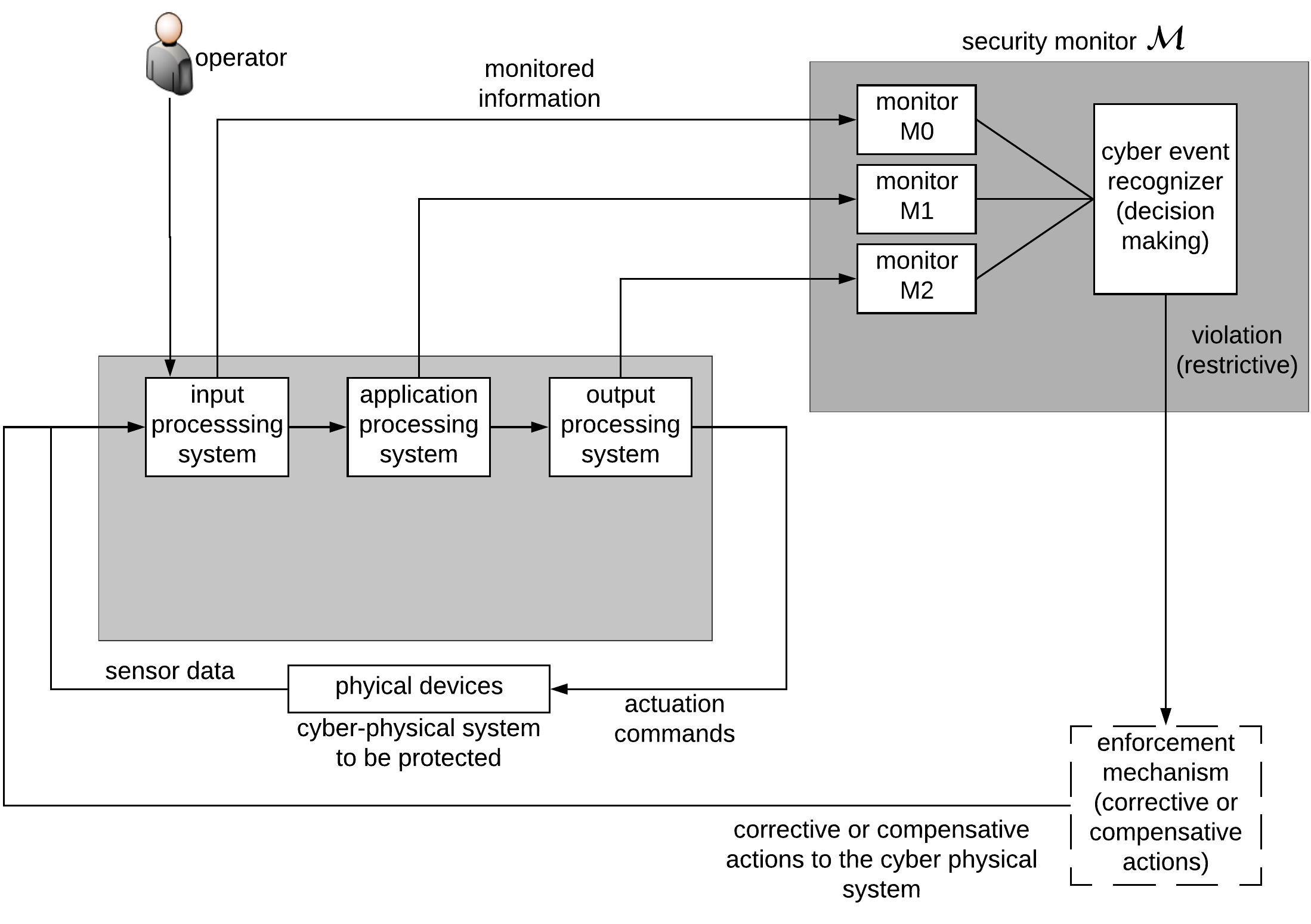}
    \caption{The multiple monitor setup observes several different domains in the \textsc{cps}.}
    \label{fig:multiple_monitors}
\end{figure}

Assuming that all security violations
or faults in the system are observable,
then we can extend the definition as follows.

\begin{itemize}
    \item Let \(\mathcal{P}_\text{observable}\) be a set of all observable runs which has events and conditions that are monitorable (from the monitor language \(\mathcal{M}_w\)) and acted upon by monitor \(M\) defined by \(\mathcal{M}_w\). \(\mathcal{P}_\text{observable}\) has one on one correspondence with monitor language \(\mathcal{M}_w\).
    \item Let \(\widehat{\mathcal{P}}\) be a detection predicate over the streams of data \(\mathcal{M}_w\) and \(\beta\) be the set of all safe runs in the observed data stream.
    \item Let \(\mathcal{P}_\text{safe}\) be a condition that holds true if there exists only \(\beta\) from the set of all observable runs.
    \item Let \(\alpha\) be a set of bad finite prefixes (violations over \(\mathcal{P}_\text{safe}\)) from the set of all runs.
\end{itemize}

Then, the monitor \(M\) witnesses \(\mathcal{P}_\text{observable}\) as follows:

\begin{gather*}
\widehat{\mathcal{P}}(\mathcal{P}_\text{observable}) \implies \beta \in \mathcal{P}_\text{observable} \implies \mathcal{P}_\text{observable} \in \mathcal{P}_\text{safe} \\ \therefore \mathcal{P}_\text{safe} = \text{holds}.
\end{gather*}

When the detection predicate acts 
on the observable data \(M_w\)
and the data is safe and has no security exploits,
then it implies that only the safety runs  belong to \(\mathcal{P}_\text{observable}\)
and the safety property \(\mathcal{P}_\text{safe}\) holds.

\begin{gather*}
\neg \widehat{\mathcal{P}}(\mathcal{P}_\text{observable}) \implies (\exists \alpha : \alpha \beta \notin \mathcal{P}_\text{safe}) \\
\therefore \mathcal{P}_\text{safe} = \text{rejected}.
\end{gather*}

When the detection predicate acts 
on the observable data \(M_w\)
and has detected a fault or an attack, 
then there exists at least one bad prefix \(\alpha\)
with the safety run \(\beta\).
Therefore, the safety property \(\mathcal{P}_\text{safe}\) ceases to hold.

The above definition is provisioned 
on two sufficient conditions: 

\begin{cond}
The detection predicate(s) \(\widehat{\mathcal{P}}\)
that define the safe execution
of a run must be defined
within the monitor and be under control of the monitor.
\end{cond}

\begin{cond}
External processes to the monitor cannot manipulate, gain access or view the detection predicate(s) \(\widehat{\mathcal{P}}\) 
within monitor \(M\).
\end{cond}

Therefore, the monitor \(M\)
with all the observable data \(\mathcal{M}_w\)
makes real-time safety and security assessments 
of the \textsc{cps}
using the detection predicate \(\widehat{\mathcal{P}}\).

A \textsc{cps} can have attacks 
at different levels 
of the system. 
The attacks can be 
on the inputs and outputs 
of the system such as 
on the sensors and actuators.
They can also be on the communication channel such as changes in baud rate, denial of service attacks. 
To detect attacks at multiple levels in a system
and to identify and isolate the attacks we require  multiple collaborative monitors rather
than a single monitor.

\subsection{Extending to Multiple Monitors}

Multiple collaborative monitors observe the monitorable data \(\mathcal{M}_w\)
and detect attacks occurring at different levels 
in the target system.
Furthermore, multiple monitors provide many layers 
of defense against an attack 
and faster detection and isolation of an attack.
The single monitor M, is extended 
to comprise of multiple monitors and is denoted as: \(\mathcal{M} = \left(M_1, M_2, \dots, M_n\right)\) (Figure~\ref{fig:multiple_monitors}).

The natural organization 
of an embedded \textsc{cps} suggests a specific grouping 
of the monitors. 
For example, if certain events in the \textsc{cps} are checked
by monitor \(M_1\) and there are prerequisite events 
for the monitor \(M_1\)
to make decisions which are checked by monitor \(M_2\),
then the grouping of these two monitors follow a specific configuration. 
In general, monitors can be grouped as sequential, parallel, associative,
and complimentary configurations~\cite{elks:2005}.

One way to formalize and model the multiple monitor architecture
and the system it targets is by reasoning 
in assets and their dependence relationships between these assets,
\(G = \left(\Sigma, \mathcal{M}, E\right)\),
where \(\Sigma\) the target \textsc{cps}, \(\mathcal{M}\) the multiple monitors as defined above, and \(E\) the edges between monitors, between monitors and the system, and vice versa.

\begin{figure}[!t]
    \includegraphics[width=0.5\linewidth]{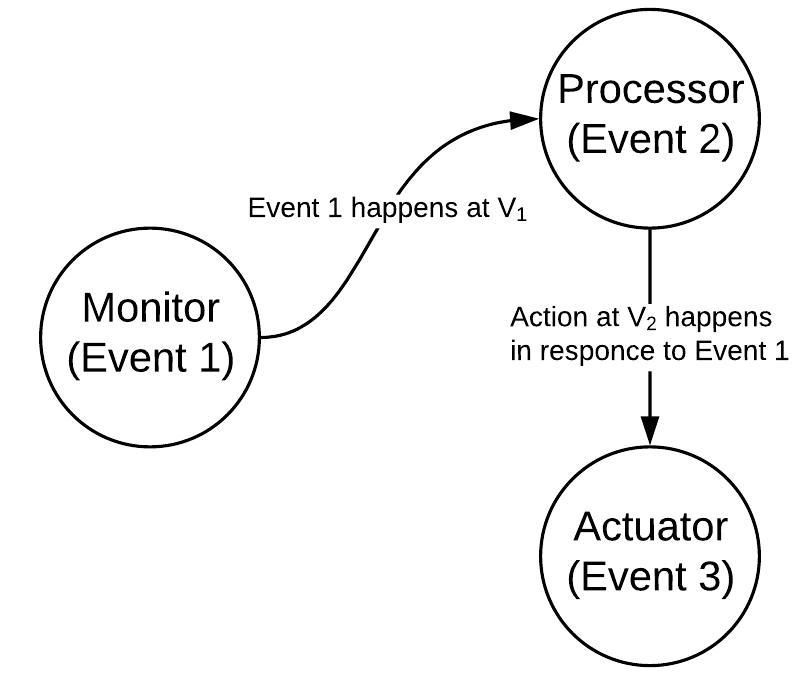}
    \caption{The graph structure allows for representing event calculus fluents as dependence relationships within the \textsc{cps} and a monitor.}
    \label{fig:graph}
\end{figure}

In summary, each monitor element \(M \in \mathcal{M}\) 
and each system element \(\sigma \in \Sigma\) form
the vertices of the graph and the interactions between
any of these elements construct the edges.
The interactions among the monitors is one
of the key elements of the architecture
and are initiated by occurrences
of events observed in the monitored stream \(\mathcal{M}_w\)
from the \textsc{cps}, \(\Sigma\).
To further formalize these interactions
such that we can implement the detection predicates---a set
of safety and security conditions/guards in the monitor---we use event calculus~\cite{spanoudakis:2007,talcott:2008,franceschet:2000} (Figure~\ref{fig:graph}).

\subsection{Expressing Monitoring Events}
Event calculus is a temporal language that represents properties
of a system~\cite{shanahan:1999}. Based on those properties it can, further,
represent the potential consequences of an action over the system.
More generally, event calculus logically describes 
(i) the properties that hold true in a system and
(ii) when they hold true.
This is determined based on two conditions: 
(a) ``What happens and when?'': the events that occur in a system and the time at which they occur. 
(b) ``What do actions do?'': How the action taken on a system can change the state of the fluents of the system.
Here a fluent is analogous 
to a variable whose value can change over time.
The following are standard predicates in event calculus:
\begin{itemize}
    \item \(\textsf{Happens}(\alpha, t)\) means that an action \(\alpha\) happens at time \(t\).
    \item \(\textsf{Initiates}(\alpha, f, t)\) means that an action \(\alpha\) occurs at time \(t\) and a fluent \(f\) starts to hold true. Fluent \(f\) that holds true at the start of an operation is shown by the predicate \(\textsf{InitiallyP}(f)\). 
    \item \(\textsf{Terminates}(\alpha, f, t)\), means the termination of a fluent which signifies that fluent \(f\) stops being true after an action \(\alpha\) occurs at time \(t\). 
    \item \(\textsf{HoldsAt}(f, t)\) shows that the fluent \(f\) holds at time \(t\). 
    \item \(\textsf{Clipped}(t_1, f, t_2)\) indicates that the fluent \(f\) holds true between time period \(t_1\) and \(t_2\). 
\end{itemize}

Events are observable states
in the finite prefix
of a stream that express real time properties.
This provides a method to describe safety 
and security predicates in the context of streams. 
If some properties are violated
and the monitor detects that the system is unsafe, 
a mitigation strategy must commence.
An example of such a strategy is to transition the system 
to a fail-safe mode.

The multiple monitor observe different kinds of phenomena
as they pertain to the safety and security status
of the \textsc{cps}.
Ontologically, the three main domains
for an exhaustive safety and security monitor are:
(1) hardware integrity, (2) information integrity,
and (3) execution integrity.

To construct such monitors using the formalism presented above,
it is necessary to put together the graph formalism; that is, the structure,
with the event calculus formalism; that is, the conditionals
in which unsafe or insecure events are detected.

To do so we label the vertices and edges with a set of event calculus predicates.
Meaning that for the vertices \(V(G)\) there is a subset
of event occurrences \(\mathcal{E} = \left(\text{Event}_1, \text{Event}_2, \dots, \text{Event}_n\right)\).
For example, for an event occurrence \(\text{Event}_1 \in \mathcal{E}\) at vertex \(M_1 \in \mathcal{M}\)
there might be some influence to some vertex \(M_2 \in \mathcal{M}\),
its response to \(M_1\) is described through some subset
of event occurrences in \(\mathcal{E}\).
Additionally, since the two vertices \(M_1, M_2\) are dependent,
an edge in \(G\) exists between them.

The implication of this is that inbound edges to any \(M \in \mathcal{M}\) provides a means 
for monitors to observe streams \(M_w\) or communicate events occurring to other monitors.
Whereas, outbound graph edges provide propagation
of monitor decisions to other monitors
or external users of confirmation
of correct operations (for example, a safety property holding).
The output edges can also instruct the \textsc{cps}
to apply some mitigation strategy in the event
that a threat or a fault is detected.

\section{Threat Model}

To understand and correctly utilize a \textsc{cps} security monitor we first need 
to understand the threat model associated with the system under evaluation.
A concrete metric in this case can be the system's attack surface---the possible entry points 
into the \textsc{cps} by an attacker. 
To do so, we need to enumerate the fundamental functions a given \textsc{cps} has to perform based 
on its expected service, how that functionality is realized in the specific implementation, 
and which parts of that implementation are observable from the outside.

Fortunately, this information is either already established
for a number of \textsc{cps} domains~\cite{khan:2017} 
or there are methodologies
for finding the threat model or vulnerability space
for any arbitrary \textsc{cps}~\cite{bakirtzis:2018, burmester:2012}.

In general, there are three taxonomic levels 
to consider when it comes to \textsc{cps} security~\cite{leccadito:2017}.
The first is the phase of creation of the vulnerability 
in the lifecycle of the \textsc{cps}; that is, it can be in the development stage, maintenance stage,
or operational stage. 
The second is the access points; that is, individual element of the attack surface. 
Third and final describes the types of attacks that can occur at each access points.

\begin{figure}[!t]
    \includegraphics[width=0.6\linewidth]{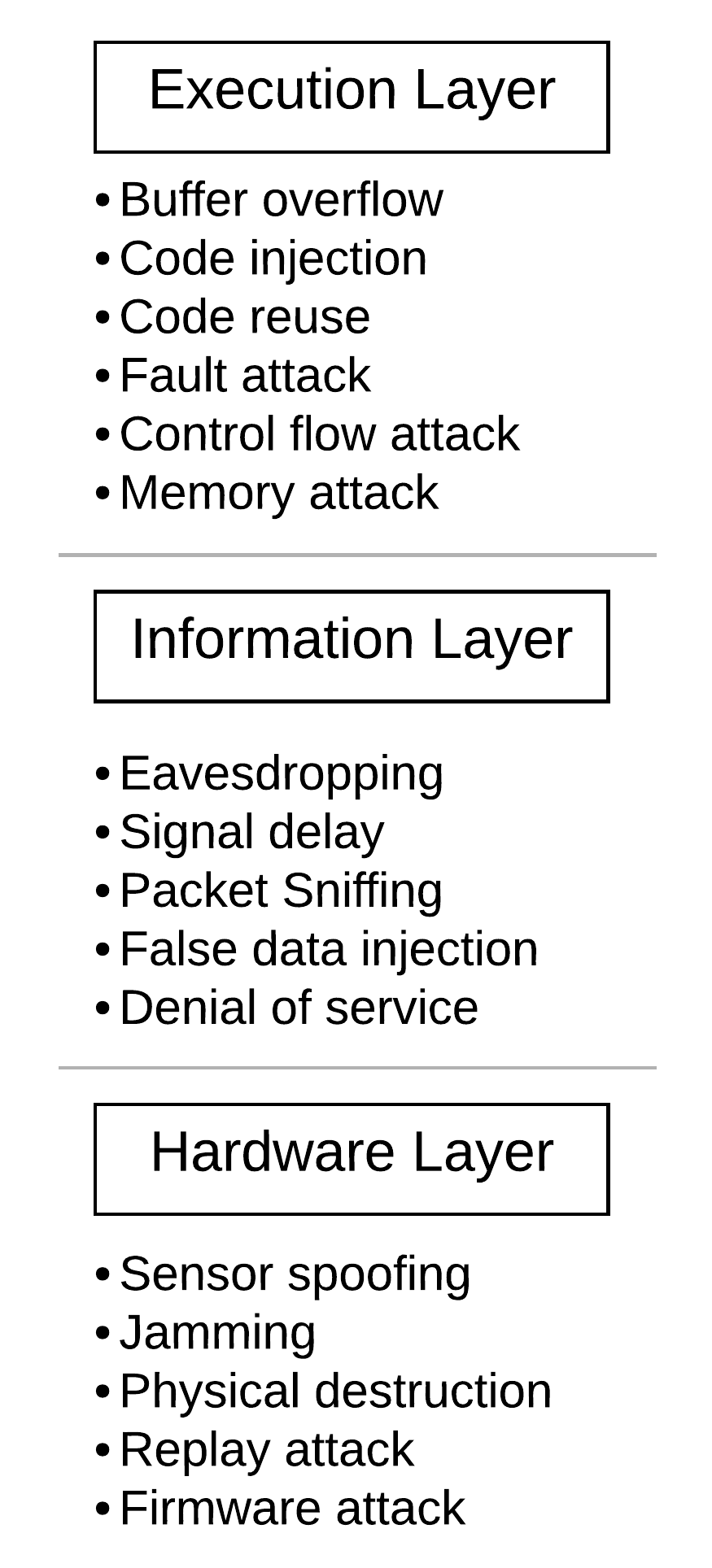}
    \caption{The threat model for an \textsc{fcs} spans three main domains: the execution layer, the information layer, and the hardware layer~\cite{leccadito:2017,lu:2015,fournaris:2017,de:2017,humayed:2017,papp:2015}.}
    \label{fig:threat_model}
\end{figure}

Therefore the threat model needs to include the hardware, information and execution layers.
These are all very appealing entrance and pivoting points for an attacker. 
Therefore monitors are needed to counteract a possible violation
at those layers. 
A thorough understanding of the vulnerabilities of the \textsc{cps} and ways of manipulating the system 
by an attacker is essential to monitor and defend the system against persistent threats (Figure~\ref{fig:threat_model}).

To evaluate our methodology and monitor design we first produce a threat model
for the \textsc{fcs}.
Specifically, the \textsc{fcs} uses an STM32F4 ARM Cotrex-M4 168 MHz microcontroller. 
Additionally, it has on-board memory components, multiple peripheral options, dedicated buses for networking,
components for the communications interfaces, and sensory systems (e.g., \textsc{gps}).
This makes it a simple but comprehensive \textsc{cps}
to produce a threat model and deploy the monitor design present in this paper.

The observable functions for the \textsc{fcs} are:
\begin{enumerate}
    \item Sensor communication: the communication between the main processor and the on-board sensors.
    \item Application software: the software programmed onto the main processor running the peripheral firmware.
\end{enumerate}

For the first function the link between the sensors and the main processor is directly monitored.
One important vulnerability that can be exploited through an attack is the violation
and consequent degradation of the on-board hardware protocols.
Specifically, an attacker can change the configuration
of the sensors's hardware peripheral communication protocol
or inject a fault into the physical signal.
Further, an attacker could exploit the firmware associated with the sensors
by injecting a malicious binary version of the sensor.
This can occur either at the supply chain by the manufacturer
or distributor. 
This does need not be an insider attack, for example,
in the case of Havex an attacker maliciously modified the manufacturers distribution website
to infect customers~\cite{vavra:2015,rrushi:2015}.

For the second functions, the design is programmed
into the main processors flash memory,
and is monitored through a debug interface such as \textsc{jtag}
or \textsc{swd}.
An attacker can exploit this interface by code injection,
where an attacker injects shell code on stack and overwrites the return address.
Specifically, code injection attempts to cause a buffer overflow; that is,
write more data than the allocated memory size which leads to the return address to be overwritten.
By doing this, the control flow of the program changes in unpredictable ways.

\section{Evaluation}
\label{sec:imp}

The implementation
of the multilayer monitor architecture contains a hardware resource integrity monitor (\textsc{hrim}), an information integrity monitor (\textsc{i2m}), 
and an execution integrity monitor (\textsc{eim}).
Thus, covering all three major domains
of vulnerability and exposure in \textsc{cps}.

\subsection{Monitoring through Event Calculus Fluents}
\label{subsec:ec}

\textbf{\textsc{hrim}.} \quad \textsc{hrim} monitors the hardware communication protocol, physical signal faults,
and configuration information. 
This monitor has no knowledge of the sensor
and receives only the bus configuration information
from the bus used by the \textsc{cps}.
It uses this information to check if the bus protocol of the sensor matches the bus configuration data received
by \textsc{i2m}.
If they match, then the sensor data is written into registers which can be accessed by \textsc{i2m}.
If it does not, then it enables a crossbar switch,
which disconnects the sensor.
To mitigate, \textsc{hrim} attempts
to reconfigure the faulty sensor
and if successful, it reconnects the sensor.

The following predefined event calculus fluents are used by \textsc{hrim}:
\begin{itemize}
    \item \textsf{sensor\_okay}, which indicates that the sensor is in expected working condition and, therefore, no fault or security violation has been found.
    \item \textsf{bus\_config\_okay}, which indicates that the communication bus is working as expected and, therefore, the communication protocol has not been tampered with.
    \item \textsf{sensor\_reconfig}, which attempts to mitigate a sensor that has been deemed to be under fault or security violation.
    \item \textsf{\textsc{hrim}\_data\_ready}, which indicates that there is no fault or security violation detected for the data received by the system hardware.
\end{itemize}

Based on the occurrence of events in the system we define the following actions:

\begin{itemize}
    \item \textsf{read\_sensor\_data}, which indicates that the sensor data is ready to be read by other subsystems.
    \item \textsf{store\_sensor\_data}, which indicates the sensor data is not faulty and is stored in data registers that could be thereby be read by other subsystems.
    \item \textsf{cross\_bar\_en}, which indicates the sensor is faulty and crossbar is enabled to disconnect the sensor.
    \item \textsf{\textsc{i2m}\_send\_InfoToDisconnect}, which makes the \textsc{i2m} to send sensor information to \textsc{hrim} when it has to disconnect the faulty sensor as \textsc{hrim} has no knowledge of the sensor.
\end{itemize}

Having predefined some baseline fluents and actions we can now define the behavior of \textsc{hrim}
through a sequential pattern of events:
\begin{enumerate}
    \item \(\textsf{InitiallyP}\left(\textsf{sensor\_okay}\right)\);
    \item \(\wedge \neg \textsf{Clipped}\left(t_i, \textsf{bus\_config\_okay}, t_n\right)\\ \wedge t_i < t < t_n \implies \textsf{HoldsAt}\left(\textsf{sensor\_okay}, t\right)\);
    \item \(\textsf{HoldsAt}\left(\textsf{sensor\_okay}, t\right)
    \implies \\
    \wedge \textsf{Happens}(\textsf{read\_sensor\_data}, t)\\
    \wedge \textsf{Happens}(\textsf{store\_sensor\_data}, t_2)\\
    \wedge \textsf{Initiates}(\textsf{store\_sensor\_data}, \textsf{\textsc{hrim}\_data\_ready}, t)\\
    \wedge t < t_2
    \);
    \item \(\neg \textsf{HoldsAt}(\textsf{sensor\_okay}, t) \implies\\
    \wedge \textsf{Happens}(\textsf{\textsc{i2m}\_send\_InfoToDisconnect} ,t)\\ 
    \wedge \textsf{Happens}(\textsf{cross\_bar\_en},t)\\
    \wedge \textsf{Initiates}(\textsf{cross\_bar\_en, sensor\_reconfig}, t)
    \).
\end{enumerate}

\textbf{\textsc{i2m}.} \quad \textsc{i2m} monitors the integrity of the data received from \textsc{hrim} by performing certain data verification actions to detect potential attacks. It also checks for subsystem inactivity and repeated values. Specifically, it waits for the data ready signal from the \textsc{hrim}, \textsf{\textsc{hrim}\_data\_ready}, and reads the sensor data from the \textsc{hrim} data registers when it is ready. If the sensor is has no detected faults or security violations and the data ready signal is not received after a certain predefined time \(t_d\), then \textsc{i2m} enables the cross bar switch and disconnects the sensor. After reading the sensor data, \textsc{i2m} additionally verifies the data by performing data integrity checks. If the sensor data passes the verification tests, the data is ready to be used by the \textsc{cps}.

The following are the predefined fluents for \textsc{i2m}:
\begin{itemize}
    \item \textsf{sensor\_idle}, which indicates that the sensor is idle. This is the initial state of the sensor.
    \item \textsf{\textsc{i2m}\_parse\_data\_success}, which indicates that  \textsc{i2m} has performed integrity checks on the data received from \textsc{hrim} and the data has no faults or detected security violations.
    \item \textsf{\textsc{i2m}\_data\_ready}, which indicates that the data is not faulty and lets \textsc{i2m} write this data into registers to be read by other subsystems.
\end{itemize}

We extend the possible actions given the requirements of \textsc{i2m}.
\begin{itemize}
    \item \textsf{\textsc{i2m}\_read\_data}, which indicates \textsc{i2m} reads data from the registers. The data is written to the registers by \textsc{hrim}.
    \item \textsf{\textsc{i2m}\_parse\_data}, whcih indicates that \textsc{i2m} performs integrity checks on the data read from the registers.
    \item \textsf{store\_I2M\_data}, I2M stores the non-faulty data on to registers that can be read by other subsystems after data verification.
\end{itemize}

Following these definitions, the pattern of \textsc{i2m} is as follows:
\begin{figure}[!t]
    \includegraphics[width=1.0\linewidth]{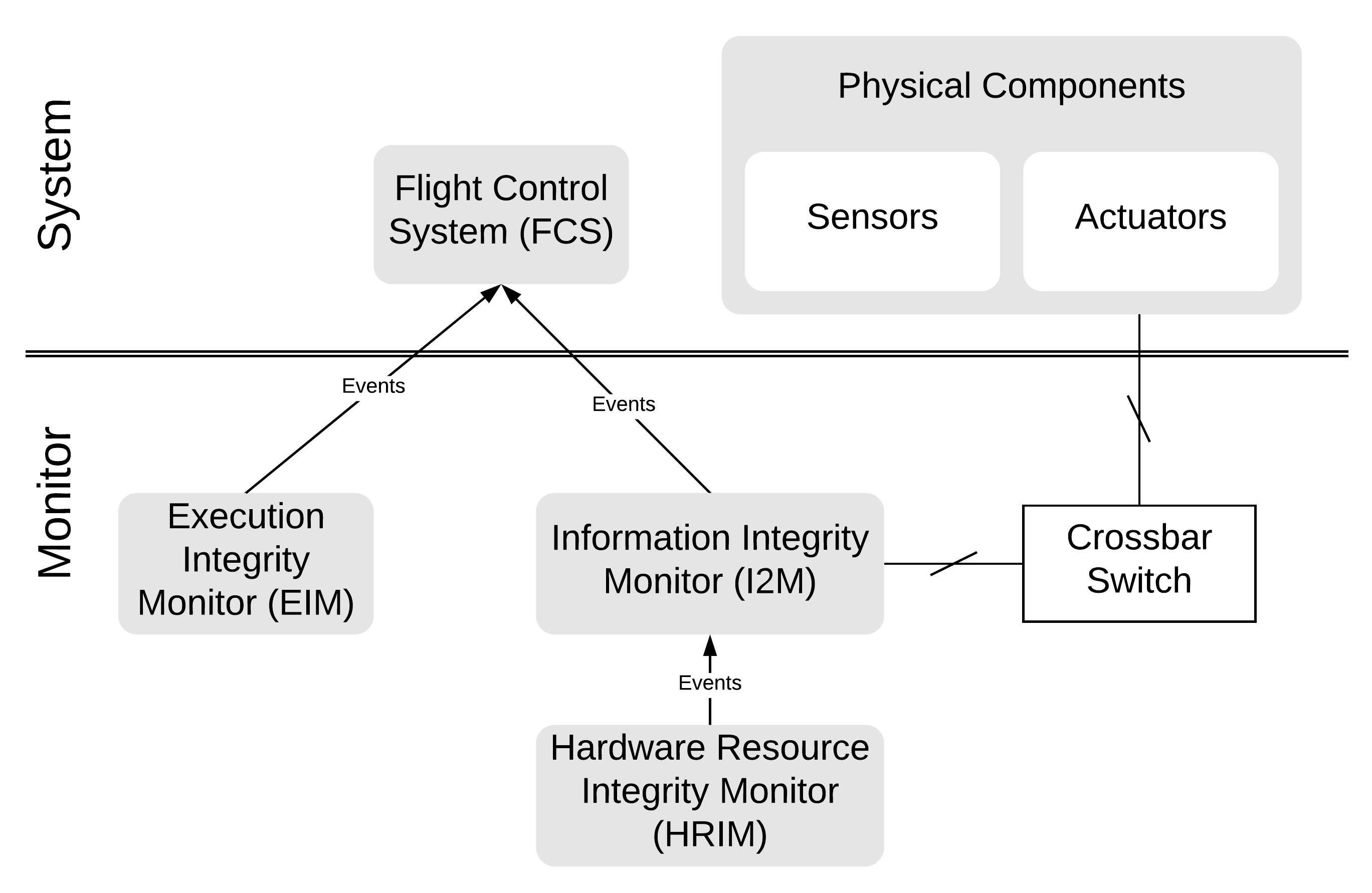}
    \caption{The implementation of the multilayer monitor architecture has a clear boundary between the system and the monitor.}
    \label{fig:architecture}
\end{figure}

\begin{enumerate}
    \item 	\(\textsf{InitiallyP}(\textsf{sensor\_idle})\);
	\item \(\textsf{Terminates}(\textsf{\textsc{i2m}\_read\_data, sensor\_idle, t})\);
	\item \(\textsf{HoldsAt}(\textsf{\textsc{hrim}\_data\_ready}, t_1) \implies\\
	\wedge \textsf{Happens}(\textsf{\textsc{i2m}\_read\_data},t_2)\\
	\wedge \textsf{Happens}(\textsf{\textsc{i2m}\_parse\_data}, t_3)\\
	\wedge t_1 < t_2 < t_3  \);
	\item \(\textsf{HoldsAt}(\textsf{\textsc{i2m}\_parse\_data\_success}, t) \implies\\
	\wedge \textsf{Happens}(\textsf{store\_\textsc{i2m}\_data}, t)\);
	\item \(\textsf{Initiates}(\textsf{store\_\textsc{i2m}\_data}, \textsf{\textsc{i2m}\_data\_ready},t)\);
	\item \( \vee \neg \textsf{HoldsAt}(\textsf{\textsc{i2m}\_parse\_data\_success}, t)\\
	\vee \neg \textsf{HoldsAt}(\textsf{sensor\_okay}, t)\\ 
	\wedge \neg \textsf{HoldsAt}(\textsf{\textsc{hrim}\_data\_ready},t_d)\\
	\wedge t < t_d \implies \\
	\wedge \textsf{Happens}(\textsf{\textsc{i2m}\_send\_InfoToDisconnect}, t)\\
	\wedge \textsf{Happens}(\textsf{cross\_bar\_en}, t)\\
    \wedge \textsf{Initiates}(\textsf{cross\_bar\_en},\textsf{sensor\_reconfig}, t) \).
\end{enumerate}

\textbf{\textsc{eim}.} \quad \textsc{eim} first checks the firmware by comparing the memory against the static memory contents stored in the processor. If the firmware is acting as expected, then the monitor signals the application to execute the program. Moreover, it monitors the memory addresses during branching operations and ensures that the return addresses and memory addresses during \texttt{jump} and \texttt{call} instructions are not tampered.

The predefined fluents of \textsc{eim} are:
\begin{itemize}
    \item \textsf{firmware\_ok}, which indicates the firmware is verified. 
    \item \textsf{control\_flow\_ok}, which indicates that the control flow is verified.
\end{itemize}

The following actions are taken by \textsc{eim} depending on the occurrence of events in the system:
\begin{itemize}
\item \textsf{check\_firmware\_ok}, which checks that the firmware in in an expected state by comparing the memory to the static memory contents stored in the processor.
\item \textsf{check\_control\_flow\_ok}, which checks two things, (1) the control flow of the program and (2) potential tampering with return addresses during \texttt{jump} and \texttt{call} instructions.
\item \textsf{execute\_program}, which checks for control flow of the design---this happens only after verifying the correct operation of the firmware and memory of the \textsc{cps}. 
\item \textsf{fail\_safe}, which forces the \textsc{cps} to jump the execution to a failsafe part of the program in the event of a detected fault or security violation.
\end{itemize}

The patterns of events occurring in the system that are monitored by \textsc{eim} are:
\begin{enumerate}
    \item \(\textsf{Happens}(\textsf{check\_firmware\_ok}, t) \implies \\ \textsf{HoldsAt}(\textsf{firmware\_ok}, t)\);  
	\item \(\textsf{Initiates}(\textsf{execute\_program},\textsf{firmware\_ok}, t)\);
	\item \(\textsf{Happens}(\textsf{check\_control\_flow\_ok}, t) \implies\\
	\textsf{HoldsAt}(\textsf{control\_flow\_ok},t)\);  
	\item \(\textsf{Initiates}(\textsf{fail\_safe},\neg \textsf{firmware\_ok} \wedge \neg \textsf{control\_flow\_ok}, t)\).
\end{enumerate}
\begin{figure}[!t]
    \includegraphics[width=1.0\linewidth]{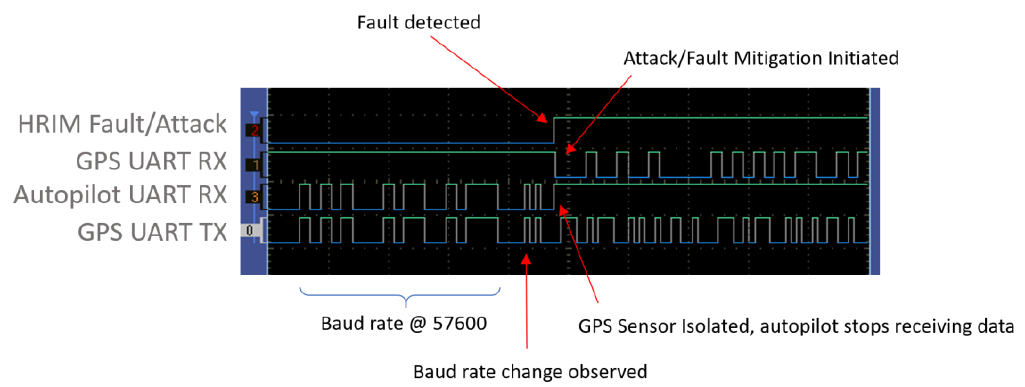}
    \caption{The \textsc{hrim} monitor can detect a baud rate change.}
    \label{fig:hrim_output}
\end{figure}
\subsection{Implementation}

The \textsc{hrim} and \textsc{i2m} monitors were implemented 
on a Nexys 4 \textsc{ddr} which is a ready to use development platform based 
on Xilinx Artix-7 XC7A100T \textsc{fpga}. 
An \textsc{fpga} board has customizable features, simulation capabilities, observability
and testability 
that makes it advantageous to implement \textsc{hrim} and \textsc{i2m}~\cite{leccadito:2017}.
Instead, the \textsc{eim} monitor is implemented on a processor---a more flexible platform
for the complex detection techniques it implements.
Specifically, we implemented \textsc{eim} on a Raspberry Pi 2 with OpenOCD running on it. 
OpenOCD is an open source software used for hardware debugging. 
Additionally, OpenOCD has a debug adapter which supports transport protocols 
such as \textsc{jtag} and \textsc{swd}~\cite{rath:2005}.
To test the multilayer architecture presented in this paper we implement monitors 
for a custom \textsc{fcs}, which in turn is implemented on  STM32F407VGT6 \textsc{arm} cortex microprocessor~\cite{ward:2014}.

The design principles and patterns governing the monitoring systems are based
on the event calculus pattern sequences (Section~\ref{subsec:ec}).
This allows us to use a formal model to understand acceptable and unacceptable states the system might transition too.
Therefore, constructing a well-formed detection mechanism
at the different layers of the monitoring architecture.

The implementation of the multilevel monitor architecture has a clear boundary
between the system and the monitoring layers (Figure~\ref{fig:architecture}).
Specifically, two of the monitors are connected serially since they need to cooperate
when producing a decision and, consequently, a mitigative action; that is, the monitor checking for hardware integrity
and the monitor that checks information integrity.
On the other hand, the execution monitor is connected in parallel checking the actual program execution
of low-level primitives.
A crossbar switch is used to decouple the digital system
from physical actions.
This is to avoid an attacker to take advantage
of either the system or the monitors to violate a physical property.

\subsection{Example Results \& Discussion}

\subsubsection{Hardware Integrity Monitor}
To demonstrate the fault detection 
and isolation capability of \textsc{hrim}, we perform an experiment 
in which a \textsc{gps} sensor baud-rate is manipulated  during run-time.
The results of injecting a physical signal fault validates the monitor's capabilities 
for detecting both security violation and intrinsic faults
and demonstrates the isolation of sensors using the crossbar switch.
Validating this functionality shows the utility of an \textsc{fpga} implementation by detecting configuration attacks and isolating a sensor
in the event of a fault or security violations.
To capture this functionality we capture the logic analyzer trace of the \textsc{uart} signals between processor and \textsc{gps} emulator (Figure~\ref{fig:hrim_output}).
\begin{figure}[!t]
    \includegraphics[width=1.0\linewidth]{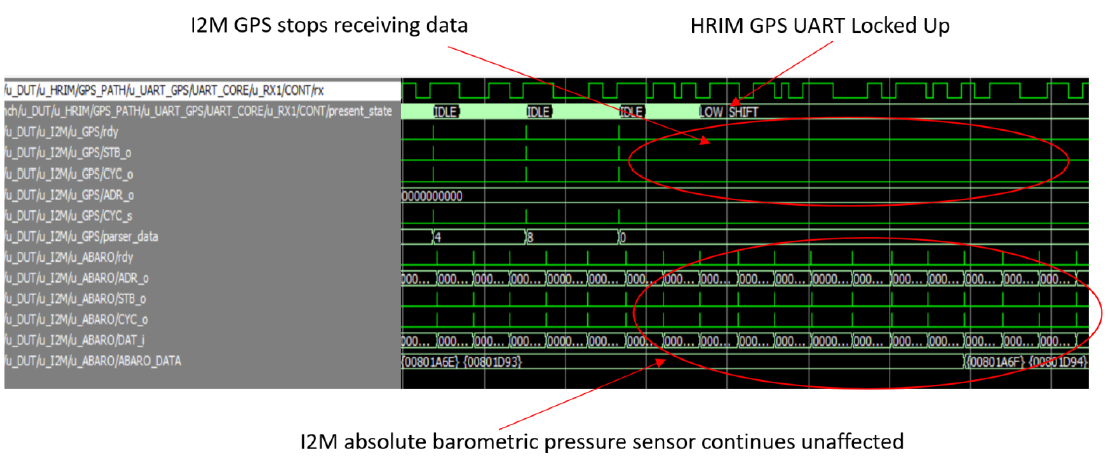}
    \caption{The \textsc{i2m} monitor can detect an \textsc{hrim gps uart} lock up.}
    \label{fig:i2m_output}
\end{figure}

The \textsc{uart} baud-rate of the autopilot and \textsc{gps} sensors are configures to 57600 Bd.
During runtime and after the initialization sequence 
of the autopilot, the baud-rate is expected to remain the same.
Instead, a fault was injected a few seconds after the autopilot booted up
to test the detection mechanism of \textsc{hrim}.
The monitor correctly detected the higher baud rate produces
by the injected fault. 
In this case the monitor forces the autopilot \textsc{uart} receive line to \textsc{idle}
and disconnects the \textsc{gps} sensor that is still sending data.
Then, the mitigation sequence is initiated by \textsc{i2m}.
Since data is being sent out of the \textsc{gps} sensor \textsc{uart} receive line
the mitigation sequence is conducted correctly.

\subsubsection{Information Integrity Monitor}

One of the tasks performed by the \textsc{i2M} apart
from data verification is, when an attack is detected by \textsc{hrim}, the I2M resets, reconfigures, 
and reconnects the sensor to the autopilot and sends a mitigation sequence 
to the \textsc{hrim} (Figure~\ref{fig:hrim_output}). 
The \textsc{i2m} monitor also detects failure
of a monitor subsystem (Figure~\ref{fig:i2m_output}).
In this example, the \textsc{hrim gps uart} locks up 
and the \textsc{i2m} does not receive data from the sensor.
Therefore triggering a fault since it expects data to be received
at a specified frequency and treats it as a sensor attack due to the inactivity.
It can be seen that the operation of other sensors, in the example below, that the barometric pressure sensor is unaffected. 
When \textsc{i2m} stops receiving data due to \textsc{hrim gps uart} lock up, it detects the fault 
and enables the crossbar switch in \textsc{hrim} which will not allow for the sensor to pass through any data to the autopilot.

\subsubsection{Execution Integrity Monitor}

During the execution of a program, control is transferred
from one part of the program to another when there is branch instruction
such as a function call or jump instruction.
To ensure that there is no return address tampering attack,
the \textsc{eim} monitors the return address during branching.
If the return address is verified
to be consistent
and, therefore, not manipulated, then \textsc{eim} concludes
that there has been no attack that leads to illegal control flow.

To achieve this, the \textsc{eim} monitor lets the application start executing
only after the checking the firmware.
Once the firmware is verified, the \textsc{eim} observes the address locations
whenever there is a branching operation.
As an example for execution monitoring, \textsc{eim} monitors the return address
to the \texttt{main} program
from a processor initialization function \texttt{mcu\_init} whenever it has finished executing
and, also, monitors the memory contents when the \texttt{main} program calls the \texttt{mcu\_init}
initialization function.

\begin{figure}[!t]
    \includegraphics[width=1.0\linewidth]{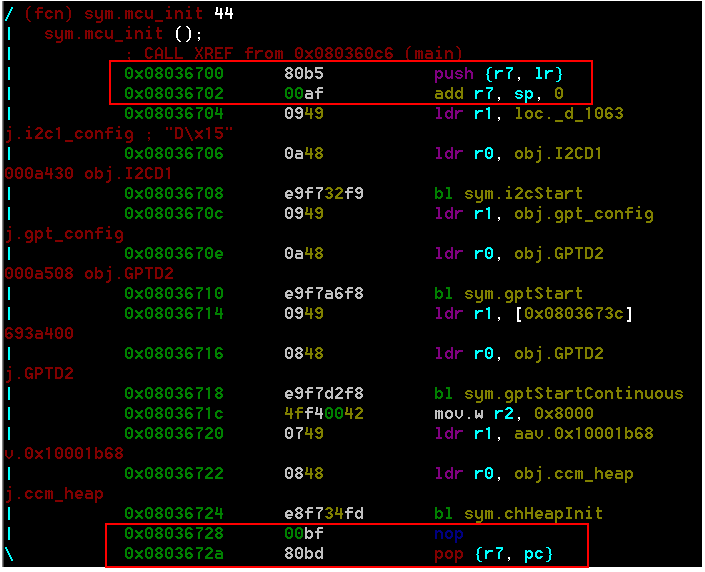}
    \caption{The disassembled code for \texttt{mcu\_init}.}
    \label{fig:mcu_init}
\end{figure}

To show the operation of \textsc{eim} we disassembled the \textsc{fcs} code
using Radare2---an open-source disassember~\cite{pancake} (Figure~\ref{fig:mcu_init}).
We visualized the information by using the \texttt{dot} files produced
by Radare2 using graphviz~\cite{ellson:2004}.

\begin{figure}[!t]
    \includegraphics[width=1.0\linewidth]{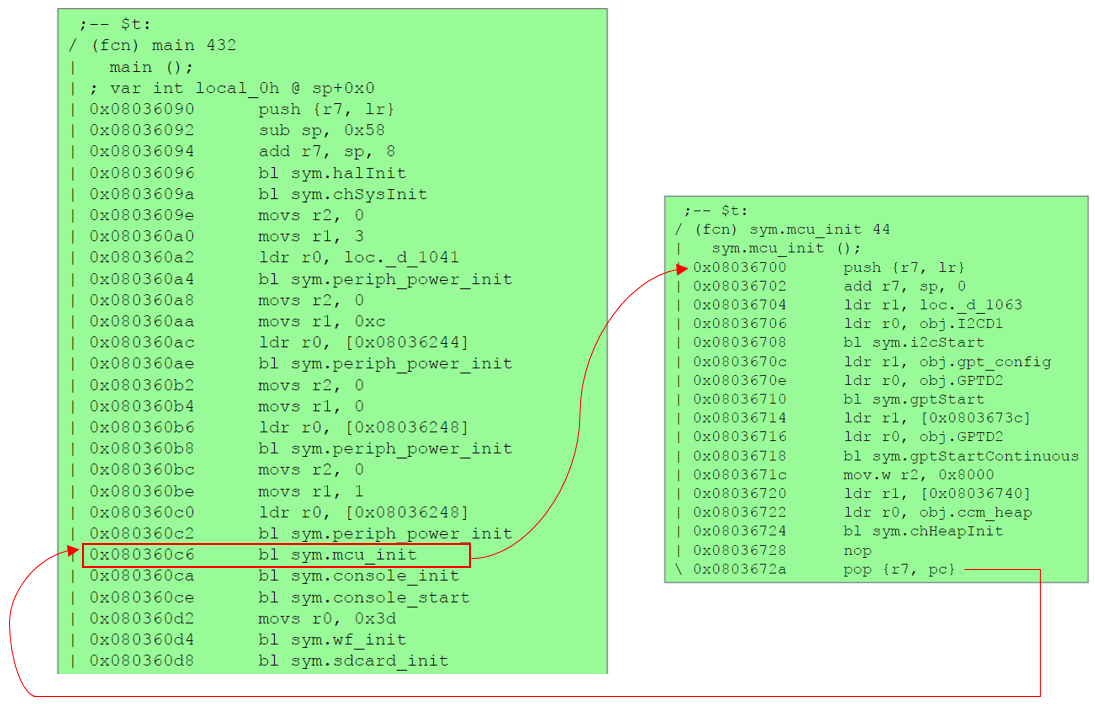}
    \caption{The control flow from the \texttt{main} function of the \textsc{fcs} to \texttt{mcu\_init} initialization function.}
    \label{fig:control_flow}
\end{figure}

\emph{Working principle.} \quad The binary values of the branching instructions are stored in the monitor and compared at runtime.
The \texttt{mcu\_init} function is called by main (Figure~\ref{fig:control_flow}). The branching instruction from main and the return instruction from \texttt{mcu\_init} are compared with the copy of data stored in the monitor at runtime.

\begin{figure}[!t]
    \includegraphics[width=1.0\linewidth]{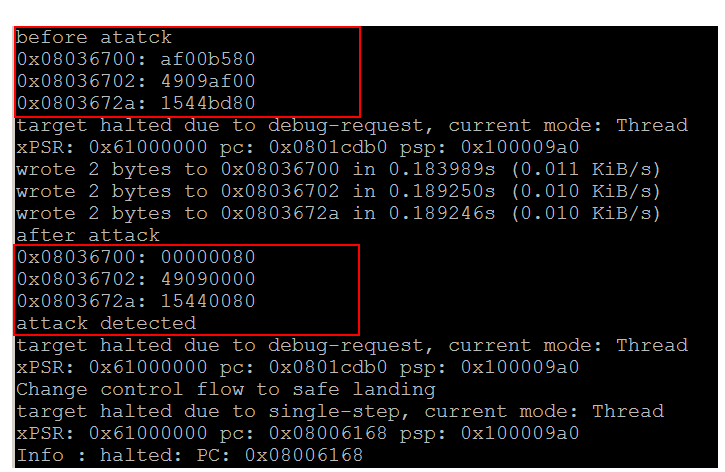}
    \caption{Output of the execution monitor using OpenOCD when the program was run while simulating an attack event. Before attack, the monitor memory matched the FCS memory. After attack, the monitor detects change in contents of FCS memory.}
    \label{fig:openocd_output}
\end{figure}

If there is an attack on the memory locations involving the branching instructions, the values 
at the memory locations would not match with the data stored 
in the monitor and it would thereby alter the control flow of the program. 
The output of the execution monitor is captured using OpenOCD when the program was run while simulating an attack event (Figure~\ref{fig:openocd_output}).
If the monitor detects the deviation from the normal branching operation, it sends the \textsc{fcs} 
to a failsafe state by changing the control flow to a safe landing function at the location \texttt{0x08006168}.

In summary, by monitoring the address locations 
at run time and comparing it with the control flow data, we are able 
to identify attacks on the system memory during branching operations.
OpenOCD was used as a prototyping platform to test non intrusive monitoring of execution conditions and events in the \textsc{fcs}.  Radare2 was used to disassemble, analyze and get a visual graphs of the functions that can be monitored using OpenOCD. Grapviz compatible \texttt{dot} file is generated from Radare2 to obtain call graphs for the functions to monitor. However, the execution of the full flight controller has thousands of functions and Graphviz could not generate a full call graphs. Since we only needed the \texttt{main} program and the function \texttt{mcu\_init} the framework could obtain the data requirements.

\section{Conclusion}

We formulate a multilevel monitor comprising of hardware, information, execution monitoring 
using a graph of assets where the dependence conditions are defined using event calculus. 
Event calculus provides a semantic foundation 
for the design of a multiple monitor architecture. 
Additionally, representing the system in a formal language helps understand the design decisions made 
at both the level of individual subsystems in a \textsc{cps}
and also globally at the system level.
Further, the use of graphs is amenable to implement control flow
with edges based on events and event calculus conditions applied to the vertices 
that can be mapped to security monitors.

To evaluate the proposed monitor architecture, we implement the hardware and information monitors on an \textsc{fpga}
and the execution monitoring on a processor. 
We have multiple monitors to watch different layers of a \textsc{cps}. 
Multiple monitors help detect the attack faster and prevent it from affecting the rest of the system. 

In the future, we plan to implement a mission monitor to observe the overall functionality
of the system and check for conditions specific to the application
at runtime and ensure that they hold true to further enhance the security of the \textsc{cps}.  Furthermore, in addition to integrating monitors across different levels, it is also possible to have multiple monitors within each level which are horizontally integrated.
Furthermore, another orientation for the organization of monitors is assume-guarantee compositions, which can be explored as an extension to this work.



\bibliographystyle{ACM-Reference-Format}
\bibliography{manuscript}

\end{document}